# Graphene photodetector integrated on a photonic crystal defect waveguide

Simone Schuler[1,*], Daniel Schall[2], Daniel Neumaier[2], Benedikt Schwarz[3], Kenji Watanabe[4], Takashi Taniguchi[4] and Thomas Mueller[1,*]

[1] *Vienna University of Technology, Institute of Photonics, Gußhausstraße 27-29, 1040 Vienna, Austria*
[2] *AMO GmbH, Otto-Blumenthal-Straße 25, 52074 Aachen, Germany*
[3] *Vienna University of Technology, Institute of Solid State Electronics, Floragasse 7, 1040 Vienna, Austria*
[4] *National Institute for Materials Science, 1-1 Namiki, Tsukuba, 305-0044 Japan*

*Corresponding authors: simone.schuler@tuwien.ac.at, thomas.mueller@tuwien.ac.at

**We present a graphene photodetector for telecom applications based on a silicon photonic crystal defect waveguide. The photonic structure is used to confine the propagating light in a narrow region in the graphene layer to enhance light-matter interaction. Additionally, it is utilized as split-gate electrode to create a pn-junction in the vicinity of the optical absorption region. The photonic crystal defect waveguide allows for optimal photo-thermoelectric conversion of the occurring temperature profile in graphene into a photovoltage due to additional silicon slabs on both sides of the waveguide, enhancing the device response as compared to a conventional slot waveguide design. A photoresponsivity of 4.7 V/W and a (setup-limited) electrical bandwidth of 18 GHz are achieved. Under a moderate bias of 0.4 V we obtain a photoconductive responsivity of 0.17 A/W.**

Graphene has already proven to be an attractive material in photonics due to its ultra-broadband light absorption [1,2,3], electrical tunability of the optical response [4] and high carrier mobility or ultra-fast optoelectronic response, in general [5]. Furthermore, graphene can be integrated into virtually any optical platform due to its two-dimensional structure and van der Waals rather than covalent bonding with the substrate [6,7]. The integration of graphene photodetectors with silicon waveguides has been realized using different device concepts [8,9,10] and the performance has been steadily improved by using more advanced designs [11,12,13,14]. Recent work has shown promising results for high-speed photodetectors with record electrical bandwidths up to 130 GHz [15], setting a new benchmark for waveguide-integrated photodetection. Beside detectors, graphene modulators have been successfully integrated with optical waveguides [16,17,18,19,20] and a combination of a detector and a modulator within a single device was realized [21]. Much progress has also been made in large-area and high-quality graphene growth, allowing the integration of graphene on wafer-scale [22,23,24,25]. Nevertheless, for graphene photodetectors there remains a gap in terms of photoresponsivity to state-of-the art technologies based on Ge or III-V semiconductors [26,27,28,29]. A good understanding of the underlying conversion mechanisms [30] and, based on this, a careful device design could potentially allow to overcome this gap.

Here, we present a photodetector based on the integration of graphene with a silicon photonic crystal (PhC) defect waveguide. The device response relies on the photo-thermoelectric (PTE) effect, which is the dominant conversion mechanism in graphene at zero bias [31,32,33]. The large optical phonon energy in graphene (~0.2 eV) [34] and the low scattering rate of acoustic phonons [35,36] give rise to an increased temperature of photo-excited carriers well above the lattice temperature. A photovoltage is generated from the photo-excited carriers, if the Seebeck coefficient, controlled by the doping level, as well as the temperature profile in the graphene vary. As the response is generated from hot electrons, large electrical bandwidths can be achieved [37]. Another advantage is that there is no need for an external bias voltage which allows for zero-dark current operation.

## RESULTS AND DISCUSSION

A conceptual illustration of our photodetector is shown in Fig. 1a. The detector is realized on a PhC defect waveguide [38,39,40,41,42] where a line-defect is introduced by an air slot [43,44]. The photonic structure has twofold function: First, the light is guided in the slot, which leads to enhanced light matter interaction and thus an increase of the carrier's temperature $T_e$ in graphene. Second, the slot divides the photonic structure into two electrically insulated parts. These are utilized to electrically control the doping in the graphene layer to both sides of the optical absorption region in order to generate a pn-junction. Due to the lateral light confinement in the PhC structure, the silicon area on both sides of the waveguide can be extended, allowing to control the Seebeck coefficients in the entire graphene layer. Hence, by fully exploiting the temperature profile in graphene the photoresponse is further increased as compared to our previous work [12] in which we employed a conventional slot waveguide (Fig. 1c). Assuming same device parameters as in Ref. [12] (13 nm hBN, 30 μm device length, 2000 cm²/Vs carrier mobility), we estimate an improvement of the responsivity by a factor of ~4. A microscope image of the device is shown in Fig. 1b.

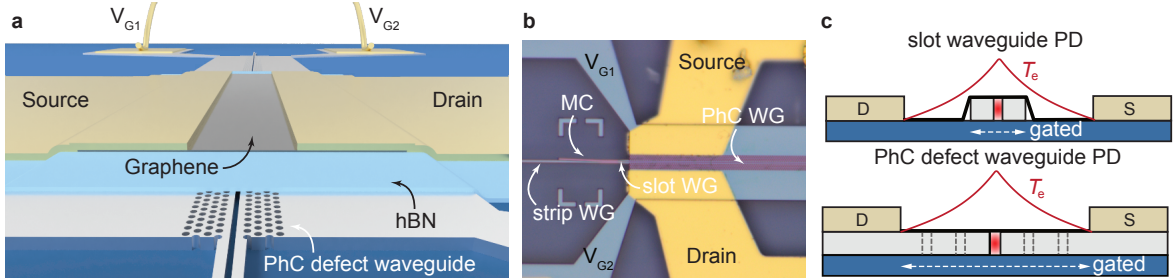

**Figure 1. (a)** Sketch of the graphene photodetector based a PhC defect waveguide. **(b)** Microscope image of a device. MC, mode converter; WG, waveguide. **(c)** Schematic illustration of a graphene slot waveguide photodetector (top; see Ref. 12) and a PhC defect waveguide detector (bottom). The electron temperature ($T_e$) in the graphene sheet is illustrated as red line. Whereas in the slot waveguide device only part of the graphene sheet is gated, the graphene in the PhC device is fully gated which allows for complete conversion of the $T_e$-gradient into a photovoltage.

The PhC was designed using *MIT photonic bands* [45] and *COMSOL Multiphysics* [46] to exhibit a photonic bandgap at the design wavelength of 1550 nm. The photonic structure was fabricated on a silicon-on-insulator (SOI) wafer with a device layer thickness of 220 nm on top of a 3 μm buried oxide layer. The silicon was weakly p-doped with a resistivity of 14-22 Ωcm. Electron-beam lithography and reactive ion etching were used to define holes with a diameter of $d = 265$ nm and a lattice period of $a = 410$ nm which build the PhC as shown in Fig. 2b. The defect waveguide was then formed by omitting one row of holes, increasing the distance of the holes to $W = 2a = 820$ nm, and introducing an air slot with a width of $w_{Slot} = 73$ nm. The fabricated PhC defect waveguides have a length of $L_{PhC} = 100$ μm. On each side of the slot, four rows of holes are introduced followed by a silicon slab. The dispersion relation of the PhC is shown in Fig. 2a. The white area indicates the TE photonic bandgap, which is defined by the modes of the crystal (dark blue). The SiO$_2$ light cone is shown in light blue color. The line-defect allows for states inside the photonic gap, marked with green and light green lines. The corresponding electric field profiles for both defect modes are shown in Fig. 2b. The slot mode (M1) is the preferred mode, due to the large field overlap with the incident mode from the slot waveguide.

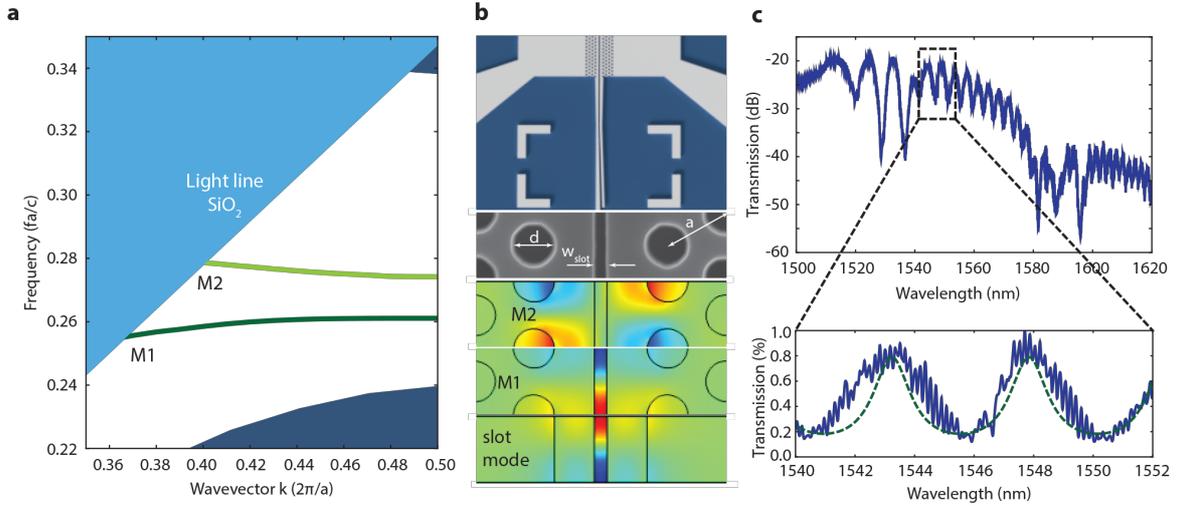

**Figure 2. (a)** Dispersion relation of the PhC waveguide. M1 and M2 are the even and odd TE modes that occur in the photonic gap due to the slot defect. **(b)** From top to bottom: Optical micrograph of the mode coupler. SEM image of the photonic crystal defect waveguide with $d = 265$ nm, $a = 410$ nm and $w_{Slot} = 73$ nm. Simulated field profiles of the modes in the PhC (M1, M2) and the slot waveguide, respectively. **(c)** Optical transmission of the photonic structure (without graphene). Oscillations are visible, which indicate the appearance of mode mismatch at the coupling interface. The enlargement shows the linear transmission (blue) in the range of the design wavelength. A Fabry-Perot model was used to model the oscillations and to extract the coupling loss (dashed green line).

In order to characterize the photonic structure, we coupled TE-polarized light from a tunable diode laser via a polarization maintaining fiber into the chip and recorded the transmitted power, while tuning the laser wavelength. Therefore, the light was first coupled with a grating coupler into a strip waveguide with a width of $d_\text{Strip}$ = 500 nm, followed by a 10 µm-long mode converter (MC) that adiabatically transfers the light into a conventional slot waveguide, and finally into the PhC defect waveguide [47]. The PhC defect waveguide confines the light to subwavelength dimension and enables the electrically separation of the two silicon parts, which are further used as dual-gate electrodes. At the end of the PhC defect waveguide we attached the same coupling scheme in order to conduct optical transmission measurements, which allowed an estimation of the coupled power. The measured transmission is depicted in Fig. 2c. The PhC slot mode (M1) is the preferred mode, due to the large field overlap with the incident mode from the slot waveguide. Light with wavelengths up to 1580 nm is guided trough the waveguide. The transmission of light with a longer wavelength is not supported since these wavelengths cannot couple into the defect mode. The oscillations on the transmission indicate the appearance of light traveling forth- and backwards in the photonic structure. Commonly this can be attributed to a mode mismatch appearing at the coupling interface between the slot waveguide and the PhC defect waveguide, thus forming an optical cavity in the PhC [48]. In order to determine the reflectivity $R$ at the interface between PhC and slot waveguides and the coupling loss, we use a Fabry-Perot model to fit the experimental transmission data:

$$T = \frac{I_T}{I_0} = \eta^2 \frac{(1-R)^2 e^{-2\alpha_\text{PhC} L_\text{PhC}}}{(1-R\, e^{-2\alpha_\text{PhC} L_\text{PhC}})^2 + 4R\, e^{-2\alpha_\text{PhC} L_\text{PhC}} \sin^2(\pi\omega/\omega_r)}, \quad (1)$$

with the incident and transmitted optical intensities $I_0$ and $I_T$, respectively, the PhC waveguide losses $\alpha_\text{PhC}$ and the mode spacing $\omega_r$. $\eta$ describes the coupling loss that includes the grating coupler and the mode converter. Neglecting losses in the PhC waveguide due to its short length ($\alpha_\text{PhC} L_\text{PhC} \approx 0$), one can derive from Eq. (1) the fringe visibility $V = (T_\text{max} - T_\text{min})/(T_\text{max} + T_\text{min}) = 2R/(1+R^2)$, from which we determine $R$ = 0.36 at the design wavelength. In order to determine $\eta$, we use Eq. (1) to fit the transmission data with $R$ = 0.36 (green dashed line in Fig. 2c), resulting in $\eta$ = -10.65 dB.

After characterization of the photonic structure, we fabricated several photodetectors on top of the waveguide structures. Therefore, in a first step the waveguides were contacted using Ti/Au pads to apply a voltage in order to control the Seebeck coefficient by electrical gating of the two distinct regions in the device. Dry-transferred hexagonal boron nitride (hBN) with a thickness of 15 nm was used to insulate the graphene layer from the Si gate electrodes. Atomic force microscopy (AFM) was performed to ensure a clean surface before transferring the graphene layer. Graphene of proper size and thickness was prepared by mechanical exfoliation on a stack of polymers on a sacrificial Si chip. The polymer stack consisted of PAA (poly acrylic acid) and PMMA (poly methyl methacrylate acid) with a thickness chosen such that flakes could be identified by optical microscopy. To verify the monolayer thickness of the graphene, Raman spectroscopy [49] was performed before transferring the sheet on the waveguide. The polymer stack was then put into water to dissolve the PAA. After a while the PMMA film floats on top of the water and was then put onto a PDMS (poly-dimethyl-siloxane) stamp which was placed beforehand on a glass slide. The stamp was turned upside down and placed with micrometer precision onto the waveguide. In order to avoid additional placement of graphitic junks on top of the waveguide structure an aperture was defined in PMMA before the actual transfer process. The PMMA layer, which was used as a transfer media, was also used to define the metallic contacts via electron-beam lithography and evaporation of Ti/Au contacts. A microscope image of the sample is shown in Fig. 1b. The gate electrodes were wire bonded and the drain and source were contacted with an RF probe (signal-ground configuration).

In the following section we present the electro-optical characterization of a typical graphene photodetector realized on 15 nm hBN with a device length $L_{Gr} = 40$ μm, a drain-source electrode spacing of 3 μm and a charge carrier mobility of ~2400 cm²/Vs. The gate-tunability of the device was verified by varying two gate voltages $V_{G1}$ and $V_{G2}$ and recording the drain-source current as shown in Fig. 3a. Four characteristic regions can be identified: p-p, p-n, n-n and n-p, demonstrating the gate tunability of the carriers in the graphene sheet. The maximum of the resistance map occurs at the charge neutrality point.

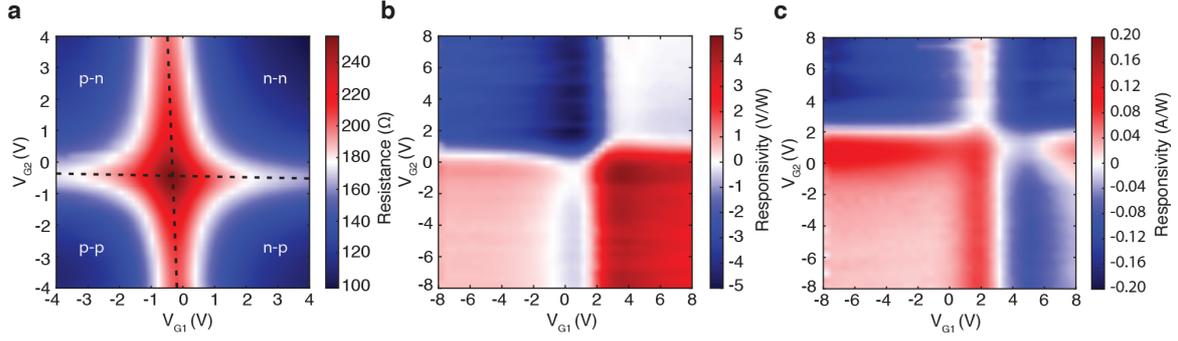

**Figure 3. (a)** Gate voltage dependent resistance map of the graphene photodetector. Four characteristic regions can be identified, p-p, p-n, n-n and n-p. The resistance peak close to zero gate voltages indicates the charge neutrality point of the graphene layer. **(b)** Measured photovoltage map of the device at zero source-drain bias. A clear six-fold pattern can be identified, indicating that the PTE effect is the dominant conversion mechanism. A maximum response of 4.7 V/W was measured. **(c)** Photocurrent map at $V_{\text{Bias}} = 0.4$ V. The response has an additional contribution due to the photoconductive effect around the charge neutrality point. A maximum response of 0.17 A/W was measured.

The optoelectronic response was measured using a tunable laser diode. Light with TE polarization at 1550 nm was coupled via the grating coupler into the device. While varying the two gate voltages $V_{\text{G1}}$ and $V_{\text{G2}}$, the photoresponse was recorded using a lock-in amplifier and a mechanical chopper. The resulting photovoltage map at zero bias is shown in Fig. 3b. The clearly visible six-fold pattern indicates that the PTE effect is the dominant conversion process. The response under bias is depicted in Fig. 3c. An additional contribution can be identified which is attributed to the photoconductive effect [50]. Given an optical power at the fiber output $P_{\text{Fiber}} = 6.5$ mW and the coupling loss $\eta$, extracted from the transmission measurement as explained above, we estimate a maximum photoresponsivity of $R_V = V_{\text{PTE}}/(\eta P_{\text{Fiber}}) = 4.7$ V/W under zero bias operation. As the PTE conversion mechanism generates a photovoltage, we give the responsivity in terms of V/W rather than the usual A/W. When operated in the photoconductive mode, our device delivers a maximum responsivity of $R_A = I_{\text{PC}}/(\eta P_{\text{Fiber}}) = 0.17$ A/W under a bias of 0.4 V.

The wavelength dependence of the device is shown in Fig. 4a ($V_{\text{G1}} = 4$ V and $V_{\text{G2}} = -2$ V). The response of the detector follows the wavelength dependence of the PhC defect waveguide. Similar to the transmission data in Fig. 2b, the photovoltage shows oscillations due to the Fabry-Perot cavity effect, but with reduced fringe contrast. This is attributed to the additional losses introduced by absorption in the graphene layer. The absorbed optical power in the graphene can be written

$$I_A = \eta I_0 \left(1 - \frac{(1-R)^2 e^{-2\alpha_{Gr}L_{Gr}} + 4R\sin^2(\pi\omega/\omega_r)}{(1-R\,e^{-2\alpha_{Gr}L_{Gr}})^2 + 4R\,e^{-2\alpha_{Gr}L_{Gr}}\sin^2(\pi\omega/\omega_r)}\right). \quad (2)$$

With $L_{Gr} = 40$ μm, $R = 0.36$ (as determined from the transmission experiment) and a graphene modal loss of $\alpha_{Gr} = 0.065$ dB/μm (obtained from mode simulations), Eq. (2) reproduces the photovoltage oscillations (inset in Fig. 4a, dashed green line).

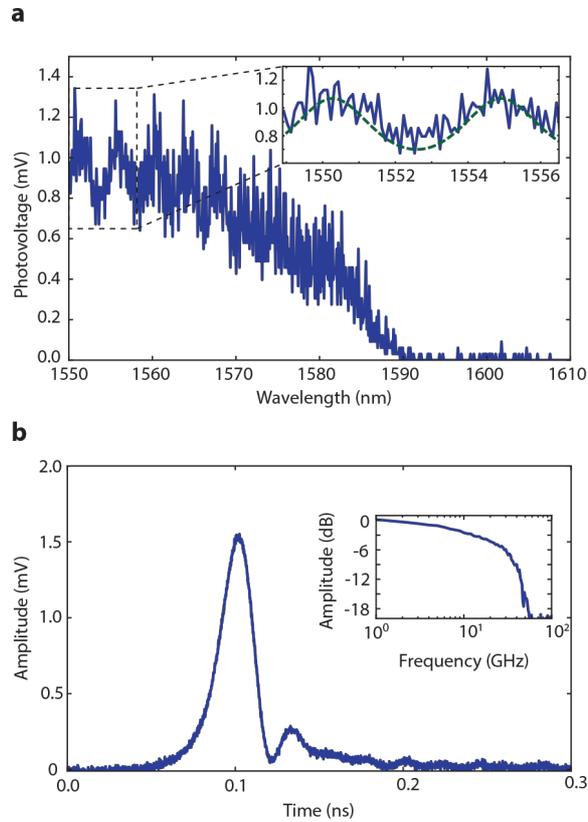

**Figure 4. (a)** Wavelength dependence of the photoresponse, which follows the response of the photonic crystal defect waveguide. **(b)** Measured impulse-response. 1 ps long pulses were coupled into the photonic structure while the response was recorded using an oscilloscope. FWHM duration of $\Delta t \approx 24$ ps is extracted from the data, which corresponds to a 3 dB-bandwidth of 18 GHz, which is the limit of the measurement setup. The inset shows the Fourier-transform of the impulse response.

To characterize the speed of our device we performed impulse-response measurements using a mode-locked erbium fiber laser which provides ~1 ps long optical pulses at a wavelength of 1550 nm. The pulses where coupled into the device via an optical fiber and the impulse-response was monitored with an oscilloscope (20 GHz bandwidth). The impulse-response is shown in Fig. 4b. The measured pulse has a full-width at half-maximum (FWHM) duration of $\Delta t \approx 24$ ps, which corresponds to a bandwidth of $f_{3dB} \approx 0.44/\Delta t \approx 18$ GHz, assuming a Gaussian pulse shape. The corresponding Fourier-transform of the impulse-response is shown in the inset of Fig. 4b and yields the same bandwidth value. The measured bandwidth is limited by the setup (mainly the oscilloscope) and the bandwidth of the graphene photodetector is expected to be higher.

## CONCLUSION

In summary, we presented a graphene photodetector based on a PhC defect waveguide relying on the PTE effect. The PhC waveguide allows for guiding the light in a confined area and furthermore to generate p- and n–doped regions in the graphene to fully exploit the PTE for the photosignal generation. Maximum responsivities of 4.7 V/W at zero bias and 0.17 A/W under a moderate bias of 0.4 V have been obtained. The photodetector has an electrical bandwidth larger than 18 GHz. The device design could be further improved by optimizing the coupling into the PhC waveguide using an improved coupling scheme.

## ACKNOWLEDGMENTS

We acknowledge financial support by the European Union (grant agreement No. 785219 Graphene Flagship) and the Austrian Science Fund FWF (START Y 539-N16).

*The authors declare no competing financial interest.*

**REFERENCS**